\documentclass[onecolumn,11pt,a4paper,aps]{revtex4}

\usepackage{colordvi}
\usepackage{graphicx,sidecap}
\usepackage{bm}
\usepackage{braket}
\usepackage{amssymb}
\usepackage{mathrsfs}
\usepackage{amsmath}
\usepackage{xcolor}
\usepackage{hyperref}
\usepackage{enumerate}
\usepackage[toc,title]{appendix}
\usepackage{verbatim}
\usepackage[margin=30pt, bf, font=footnotesize, center, justification=justified]{caption}[2004/07/16]
\hypersetup{colorlinks,bookmarksopen,bookmarksnumbered,
 citecolor=red,
 linkcolor=blue,
 pdfstartview=green,
 urlcolor=purple}

\newcommand{\ee}{\mathrm{e}}

\def\be{\begin{equation}}
\def\ee{\end{equation}}
\def\bea{\begin{eqnarray}}
\def\eea{\end{eqnarray}}

\begin{document}

\title {On the continuum limit of the entanglement Hamiltonian}
\author{ Viktor Eisler$^1$, Erik Tonni$^2$ and Ingo Peschel$^3$}

\affiliation{
$^1$Institut f\"ur Theoretische Physik, Technische Universit\"at Graz, Petersgasse 16,
A-8010 Graz, Austria\\
$^2$SISSA and INFN Sezione di Trieste, via Bonomea 265, I-34136 Trieste, Italy\\
$^3$Fachbereich Physik, Freie Universit\"at Berlin, Arnimallee 14, D-14195 Berlin, Germany
}

\begin{abstract}
  We consider the entanglement Hamiltonian for an interval in a chain of free fermions in its ground
  state and show that the lattice expression goes over into the conformal one if one includes the
  hopping to distant neighbours in the continuum limit. For an infinite chain, this can be done
  analytically for arbitrary fillings and is shown to be the consequence of the particular structure
  of the entanglement Hamiltonian, while for finite rings or temperatures the result is based on
  numerical calculations.
  
\end{abstract}
\maketitle

\section{Introduction}

It is well known that the entanglement properties of a quantum state are contained in the reduced
density matrix $\rho$ of one of the two subsystems into which the total system is divided, see e.g.
\cite{CCD09,Peschel/Eisler09}.
Writing $\rho=\exp(-\mathcal{H})/Z$, the features of $\rho$ are then encoded in the operator
$\mathcal{H}$, and for this reason it has been termed the entanglement Hamiltonian.

For the most studied quantity, the entanglement entropy, one only needs its eigenvalues, the so-called
entanglement spectrum \cite{Li/Haldane08}, which also gives information on topological features of
the system \cite{Li/Haldane08, Turner_etal10, Fidkowski10}. However, for a general understanding one also would
like to know its explicit form and structure. This will depend strongly on the system and on the way it is
partitioned. If a ladder is divided into its two legs, $\mathcal{H}$ can be proportional to the physical
Hamiltonian of the leg, see e.g. \cite{Poilblanc10,Cirac_etal11,Peschel/Chung11,Laeuchli/Schliemann12}.
If it is divided into two half-ladders, $\mathcal{H}$ corresponds to an inhomogeneous system, and results
for integrable chains \cite{Peschel/Eisler09} or from field theory \cite{Bisognano/Wichmann75} indicate
that the terms in $\mathcal{H}$ increase linearly as one moves away from the dividing point. 

In the present work, we consider such inhomogeneous entanglement Hamiltonians for critical systems
because they can be expected to show universal features. We do this in one dimension and for the case of
finite subsystems. Then there are two kinds of results.

For \emph{continuous} critical systems, conformal field theory gives the following general expression for an
interval of length $\ell$ in a larger chain, see \cite{Casini/Huerta/Myers11,Wong_etal13, Cardy/Tonni16,
Wen_etal16,Klich/Vaman/Wong17_1,Klich/Vaman/Wong17_2}
\begin{equation}
\mathcal{H} = 2\pi \ell \int_0^{\ell} dx \; \beta(x)\;T_{00}(x) \, .
\label{conf_ham}
\end{equation}
Here $T_{00}(x)$ is the energy density in the physical Hamiltonian and $\beta(x)$ is a weight factor
arising from the conformal mapping which relates the path integral for $\rho$ in the actual geometry to
that for a strip. Together with the prefactor, it can be viewed either
as a local inverse temperature \cite{Casini/Huerta/Myers11, Wong_etal13, Blanco/Casini/Myers13, Takayanagi/Ugajin13,Pretko17,Arias_etal17_1,Arias_etal17_2}
or as a local velocity of the particles described by $T_{00}$. For an
interval in an infinite chain, it is a parabola
\begin{equation}
\beta(x)=\frac{x}{\ell} \left(1-\frac{x}{\ell}\right)
\label{conf_beta}
\end{equation}
and vanishes linearly at the ends. It shares this property with the simpler case of a half-infinite
interval where $\beta(x)$ is completely linear \cite{Bisognano/Wichmann75, Bisognano/Wichmann76,Unruh76}.

For a homogeneous fermionic hopping model in its ground state, which is a \emph{discrete} critical system,
the structure of $\mathcal{H}$ is also known. Because the ground state is a simple Fermi sea,
$\mathcal{H}$ must be a free-particle lattice Hamiltonian \cite{Peschel03}
\begin{equation}
  \mathcal{H}=  \sum_{i,j=1}^N \, H_{i,j} \,c^{\dag}_i c_j \, 
\label{ent_ham}
\end{equation}
with $N$ denoting the number of sites in the subsystem. Numerical calculations
for small intervals then show that the hopping in $\mathcal{H}$ is predominantly to nearest neighbours
as in the physical Hamiltonian, but hopping to more distant neighbours also exists, although with small
amplitudes \cite{Peschel/Eisler09,Eisler/Peschel17}. For a large interval in an infinite
half-filled chain, analytical expressions for the hopping amplitudes have been found recently
\cite{Eisler/Peschel17}. From these one sees that the nearest-neighbour hopping follows
\emph{almost} the parabolic law \eqref{conf_beta}, deviating from it only slightly in the centre
of the interval, while the longer-range hoppings vary in space roughly like powers of that parabola.
The same general feature is found for finite chains \cite{Eisler/Peschel18}. Thus the discrete
$\mathcal{H}$ differs from the conformal result even if the subsystem is large and the lattice
structure should perhaps not matter. This is somewhat surprising and also intriguing,
because an operator $\mathcal{T}$ exists which commutes with $\mathcal{H}$ and \emph{does} have the
(discretized) conformal form, see \cite{Eisler/Peschel17,Eisler/Peschel18,Slepian78,Peschel04,Eisler/Peschel13}.

The way out of this discrepancy has already been indicated in \cite{Arias_etal17_1}, namely one
should include the longer-range hopping when taking the continuum limit and thereby obtain an
effective $\beta(x)$. Doing that numerically, the authors obtained a rather good approximate parabola.
In this communication, we want to go a step further and show analytically that the conformal
parabola results. The key ingredient is a relation which expresses the matrix $H$ in \eqref{ent_ham}
as a power series of the tridiagonal matrix $T$ commuting with it. While this relation leads
to relatively complicated matrix elements for $H$, it turns out in the end that for the continuum
limit at half filling only the lowest power of $T$ contributes and that this can be understood
from the particular structure of $T$. For general filling, the mechanism is not quite as simple, but the
result is the same.

In the following Section\;\ref{sec:cont-limit} we revisit the continuum limit for homogeneous and inhomogeneous
hopping models and derive a general expression for the resulting $\mathcal{H}$ and the
quantity $\beta(x)$. In Section\;\ref{sec:half-filling}, we evaluate $\beta(x)$ for the entanglement Hamiltonian
one encounters in an infinite half-filled chain and give an interpretation of the mechanism.
We also present numerical results for finite chains and finite temperatures. In Section\;\ref{sec:arb-filling}, we
consider chains with arbitrary filling and show that the same $\beta(x)$ results.
Section\;\ref{sec:conclusions} contains our conclusions and three appendices give details on the commuting operator, the
summations needed for general filling and on higher derivatives in the continuum limit.

\section{Continuum limit of hopping models}
\label{sec:cont-limit}

In this section, we consider hopping models with a bipartite structure where the hopping only takes
place between even and odd lattice sites. This is the situation for the entanglement Hamiltonian
if the system is half-filled. Setting $H_{i,i+2p+1}= -t_{2p+1}(i+p+1/2)$ in \eqref{ent_ham},
where $i+p+1/2$ is the midpoint between initial and final site, the Hamiltonian then takes
the form
\begin{equation}
\mathcal{H}= -\sum_{i} \sum_{p \ge 0} \, t_{2p+1}(i+p+1/2)\left( c^{\dag}_i c_{i+2p+1} + c ^{\dag}_{i+2p+1} c_i \right) \, .
\label{hopham1}
\end{equation}

\subsection{Homogeneous case}

It is instructive to consider first the case where the $t_{2p+1}$ do not depend on the position.
The corresponding Hamiltonian will be denoted by $\mathcal{H}_h$.
For a ring with $N$ sites and lattice spacing $s$, a Fourier transformation then gives
\begin{equation}
  \mathcal{H}_h= \sum_q \,\nu_q \, c^{\dag}_q c_q \, , 
\hspace{.4cm}
  \nu_q=  -\sum_{p \ge 0} t_{2p+1} \, 2\cos[(2p+1)qs]  \, , 
\hspace{.5cm}
  q=\frac{2\pi}{Ns}\,k\, ,\;\;\;\; k=0,\pm1,\pm2,\dots
\label{hopham2}
\end{equation}
and $\mathcal{H}_h$ has a half-filled ground state with Fermi momentum $q_F = \pm \pi/2s$ if $t_1$
dominates. To obtain its continuum limit, one first shifts the right and left parts of the dispersion
relation in such a way that both Fermi points lie at the origin. This is done by introducing new Fermi
operators for even and odd sites (see e.g. \cite{Fradkin13})
\begin{equation}
  c_{2n}= \textrm{i}^{2n}a_n\, , \quad  c_{2n+1}= \textrm{i}^{2n+1}b_n   \,.                                                   
\label{fermiop1}
\end{equation}
Then $\mathcal{H}_h$ contains only mixed terms $a^{\dag}b$ and $b^{\dag}a$ and is diagonalized by writing
the Fourier-transformed quantities as 
\begin{equation}
  a_q= \frac {1}{\sqrt 2} \, e^{-\textrm{i}qs/2} (\alpha_q+\beta_q) \, , \quad 
  b_q= \frac {1}{\sqrt 2} \, e^{\textrm{i}qs/2} (\alpha_q-\beta_q)
\label{fermiop2}
\end{equation}
which leads to
\begin{equation}
  \mathcal{H}_h= \sum_q \,   \omega_q \left( \alpha^{\dag}_q  \alpha_q -  \beta^{\dag}_q  \beta_q \right) \,,
  \quad  \quad \omega_q = \sum_{p \ge 0}(-1)^p \, t_{2p+1} \, 2\sin[(2p+1)qs] \,.
\label{hopham3}
\end{equation}
The Brillouin zone is now limited by $\pm\pi/2s$ and the operators $\alpha_q$  $(\beta_q)$ describe right (left)
moving particles with energy $\omega_q$ and velocity  
\begin{equation}
  v_q = \frac{d\omega_q}{dq} = \sum_{p \ge 0}(-1)^p \, (2p+1)s \, t_{2p+1} \, 2\cos[(2p+1)qs]\,.
\label{velocity}
\end{equation}
For $q=0$, this becomes the Fermi velocity
\begin{equation}
  v_F = 2s \,\sum_{p \ge 0}(-1)^p \, (2p+1)\, t_{2p+1} 
\label{fermi_velocity}
\end{equation}
where the hopping amplitudes are multiplied by the corresponding hopping distances. The factor
$(-1)^p$ is best understood if one notes that the same result is obtained if one works in the initial
formulation \eqref{hopham2} and differentiates $\nu_q$ at $q=\pi/2s$. Then $(-1)^p$ appears because the
slopes of the functions $\cos[(2p+1)qs]$ at the Fermi points alternate with $p$. 

The form \eqref{hopham3} for small $q$ where $\omega_q=v_Fq$ is the continuum limit of $\mathcal{H}_h$
in momentum space. In real space, one can obtain it directly by introducing field operators
$\psi_1(x)$ and $\psi_2(x)$ for $a_n$ and $b_n$ which are attached to the points of the original
lattice, i.e.
\begin{equation}
  a_n \rightarrow \sqrt{2s} \, \psi_1(x) \, \, , 
  \qquad  b_n \rightarrow \sqrt{2s} \, \psi_2(x+s)\, \, ,
  \qquad \sum_n \rightarrow \, \int \frac{dx}{2s}\,.
\label{fermi_cont1}
\end{equation}
Here the lattice constant $2s$ of the sublattices has been used.
In the limit $s \rightarrow 0$, one can expand the quantities for shifted sites as
\begin{equation}
  a_{n+r} \rightarrow \sqrt{2s} \, \psi_1(x+2sr) \simeq \sqrt{2s}\,\big(\psi_1(x) + 2sr \,\psi'_1(x) + \dots \,\big)
\label{expansion}
\end{equation}
and similarly for $b_{n+r}$. This  leads to an expression for $\mathcal{H}_h$ where only terms with one
derivative remain. Either by shifting summation indices or by partial integration, this derivative can be
brought into the second place such that
\begin{equation}
  \mathcal{H}_h= -\, \textrm{i}\, v_F \int_0^{\ell} dx  \left( \psi^{\dag}_1(x) \psi'_2(x)+
              \psi^{\dag}_2(x) \psi'_1 (x)\right)
\label{hopham_cont1}
\end{equation}
where $\ell=Ns$ and $v_F$ is given by \eqref{fermi_velocity}.
The transformation to right- and left-movers corresponding to \eqref{fermiop2} for $qs \rightarrow 0$
\begin{equation}
  \psi_1= \frac {1}{\sqrt 2} \, (\psi_{\textrm{\tiny R}}+\psi_{\textrm{\tiny L}}) \, , 
  \qquad 
  \psi_2= \frac {1}{\sqrt 2} \, (\psi_{\textrm{\tiny R}}-\psi_{\textrm{\tiny L}}) 
\label{fermi-cont2}
\end{equation}
then gives the final result
\begin{equation}
  \mathcal{H}_h
  \,=\, 
  v_F \int_0^{\ell} dx \,
  \Big( 
  \psi^{\dag}_{\textrm{\tiny R}}(x)\,
  (- \textrm{i}\,\partial_x )
  \psi_{\textrm{\tiny R}}(x)
  -
  \psi^{\dag}_{\textrm{\tiny L}}(x)\,
  (- \textrm{i}\,\partial_x )
  \psi_{\textrm{\tiny L}}(x) 
  \Big)\,.
\label{hopham_cont2}
\end{equation}

\subsection{Inhomogeneous case}

The procedure outlined above can easily be generalised to hopping amplitudes that vary slowly in space. 
To obtain the continuum limit, one writes 
\begin{equation}
  t_{2p+1}(i+p+1/2) \rightarrow t_{2p+1}(x+(p+1/2)s) \,.
\label{continuum_t}
\end{equation}
Then the hopping processes between $x$ and $x+s$ and their $(2p+1)$-th neighbours to the
right lead to the following terms in $\mathcal{H}$
\begin{eqnarray}
&&
  t_{2p+1}(x+(p+1/2)s) \left[ \psi^{\dag}_1(x)\psi_2(x+(2p+1)s)-\psi^{\dag}_2(x+(2p+1)s)\psi_1(x) \right]
  \\
  &&
 +\,t_{2p+1}(x+(p+3/2)s) \left[ \psi^{\dag}_2(x+s)\psi_1(x+(2p+2)s)-\psi^{\dag}_1(x+(2p+2)s)\psi_2(x+s) \right]\,.
   \nonumber 
\label{inhom1}
\end{eqnarray}
Expanding all quantities as in \eqref{expansion} gives 
\begin{equation}
  t_{2p+1}s \left[2p(\psi^{\dag}_1\psi_2'-\psi^{\dag \prime}_2 \psi_1)+
    (2p+2)(\psi^{\dag}_2\psi_1'-\psi^{\dag \prime}_1\psi_2)\right]
  - t_{2p+1}'s \left[\psi^{\dag}_1\psi_2 -\psi^{\dag}_2\psi_1\right]
\label{inhom2}
\end{equation}
where the argument is now $x$ everywhere. For the complete Hamiltonian, the last term can be converted
into one where $t_{2p+1}$ appears by a partial integration. The boundary contributions vanish even for an
open system since the fields and the hopping amplitudes are zero outside. Thus \eqref{inhom2} becomes
effectively
\begin{equation}
  (2p+1)\,t_{2p+1} s \left[ (\psi^{\dag}_1\psi_2' +\psi^{\dag}_2\psi_1')     
                         -(\psi^{\dag \prime}_2\psi_1+\psi^{\dag \prime}_1\psi_2) \right]\,.
\label{inhom3}
\end{equation}
The second term in the bracket, which is the hermitean conjugate of the first one, now has to be
kept as it is. The final expression therefore reads
\vspace{0.2cm}
\begin{equation}
  \mathcal{H}= \int_0^{\ell} dx \, v_F(x) \,T_{00}(x)
    \label{hopham_cont3}
\end{equation}  
where the operator of the energy density is now given by
  \begin{equation}
T_{00}(x)=
  \frac{1}{2}\left[ \,
  \psi^{\dag}_{\textrm{\tiny R}}(x) \, (- \textrm{i}\,\partial_x )\psi_{\textrm{\tiny R}}(x)
  -
 \psi^{\dag}_{\textrm{\tiny L}}(x)\,
(- \textrm{i}\,\partial_x )
 \psi_{\textrm{\tiny L}}(x) 
 + \,\mathrm{h. c.}
 \, \right]
     \label{first T00 def}
\end{equation}
and it generalizes \eqref{hopham_cont2} to a spatially varying local velocity $v_F(x)$ given by using
$t_{2p+1}(x)$ in \eqref{fermi_velocity}. This has the general conformal form \eqref{conf_ham}
with $v_F(x)$ appearing in place of $2\pi\ell\beta(x)$. In the following section, we will calculate
$v_F(x)$ explicitly for the entanglement Hamiltonian of an interval.

\section{Entanglement Hamiltonian for an interval}
\label{sec:half-filling}

With the result of the previous section, we can now determine the continuum limit of the
entanglement Hamiltonian for an interval in a half-filled chain. If the total
system is infinite and in its ground state, the calculation can be carried out analytically. 
The case of a finite chain or finite temperature will be considered by
resorting to a numerical evaluation of the sums involved.

\subsection{Infinite system}
\label{sec:half-filling infinite}

For an interval of $N$ sites in an infinite chain, the entanglement Hamiltonian can be given
in closed form for large $N$ and was reported in \cite{Eisler/Peschel17}. It is extensive and the scaled matrix
$h =-H/N$ has the representation
\begin{equation}
  h = \sum_{m \ge 0} \alpha_m \beta_m T^{2m+1}
  \label{h-T}
\end{equation}
with a symmetric matrix $T$ which commutes with $H$ and is bidiagonal for half filling
with elements $T_{i,i+1}=i/N(1-i/N)$, i.e. it describes nearest-neighbour hopping with parabolically
varying amplitudes. The coefficients in \eqref{h-T} are given by
\be
\alpha_m = \frac{1}{\sqrt{\pi}} \frac{\Gamma(m+1/2)}{\Gamma(m+1)}\;,
\qquad
\beta_m = \sqrt{\pi}\; 2^{2m}  \frac{\Gamma(2m+1/2)}{\Gamma(2m+2)}\;.
\label{ambm}
\ee
As a result, $H$ contains only hopping to the $(2p+1)$-st neighbours as in section\;\ref{sec:cont-limit}, and in a
proper scaling limit $i,N \to \infty$ with $i/N$ kept fixed, the hopping amplitudes read \cite{Eisler/Peschel17}
\be
\frac{1}{N}\, t_{2p+1}(z_p) =
  \sum_{m \ge p} \alpha_m \beta_m \binom{2m+1}{m-p} z_p^{2m+1}
\label{tp}
\ee
where the variable $z_p$ defined as
\be
z_{p}=\frac{i+p}{N}\left(1-\frac{i+p}{N}\right)
\label{zp}
\ee
is the element $T_{i+p,i+p+1}$ of $T$ and symmetric under the reflection
$i+p \to N-(i+p)$. The infinite sum in \eqref{tp} can be expressed in terms of generalized
hypergeometric functions $_3F_2$ for all values of $p$, but this is not necessary for
the following calculation.

Introducing now the lattice spacing $s$, the limit which has to be considered is $s \to 0$
and $N \to \infty$, keeping $x=is$ and $\ell=Ns$ fixed. Then for any fixed $p$ the dependence
of $z_p$ on $p$ can be treated perturbatively and introduces only a correction of order $s/\ell$.
Hence, for our purposes, we can ignore this effect and work in \eqref{tp} with the
variable $z=z_0$ or, in terms of $x$ and $\ell$ and bearing in mind \eqref{conf_beta}
\be
  z(x)= \frac{x}{\ell}\left(1-\frac{x}{\ell}\right) \equiv \beta(x) \,.
\label{z}
\ee
With this simplification, the only $p$-dependent term in $t_{2p+1}$ is the combinatorial factor and
the expression for $v_F$ can be written as follows
\be
 v_F(z) =  2 \ell \sum_{m \ge 0} \alpha_m \beta_m \, S_m \, z^{2m+1}\,, 
 \qquad
S_m = \sum_{p=0}^{m} (-1)^p \, (2p+1) \binom{2m+1}{m-p}\,.
\label{vz1}
\ee
Note that we have exchanged the sums over $m$ and $p$, and the
sum over $p$ in the definition of $S_m$ now runs only up to $m$.
In other words, the hopping $t_{2p+1}$ with $p>m$ does not contribute
at order $z^{2m+1}$. We now rewrite the sum in $S_m$ by substituting
$p \to m-p$ which leads to
\be
S_m 
= (-1)^m \sum_{p=0}^{m} (-1)^{p} (2m+1-2p) \binom{2m+1}{p}\,.
\ee
The factor multiplying the binomial coefficient can be split as
$(2m+1-p) - p$ and $S_m$ can be written as a difference of two sums
\be
S_m=(-1)^m \, (2m+1) \left[\sum_{p=0}^{m} (-1)^{p} \binom{2m}{p}
- \sum_{p=1}^{m} (-1)^{p} \binom{2m}{p-1} \right] .
\ee
The second sum can be further transformed by the substitution $p \to 2m+1-p$ into
\be
\sum_{p=1}^{m} (-1)^{p} \binom{2m}{p-1}=
-\sum_{p=m+1}^{2m} (-1)^{p} \binom{2m}{p}\,.
\ee
Now one can combine the two sums and realize that the resulting
expression is nothing else but the expansion of $(1-1)^{2m}$.
Hence one arrives at the simple result
\be
S_m = (-1)^m \, (2m+1) \sum_{p=0}^{2m} (-1)^{p} \binom{2m}{p}=
\delta_{m,0}
\label{Sm}
\ee
and in \eqref{vz1} only the $m=0$ term survives. Using $\alpha_0\beta_0=\pi$ and $z=\beta$,
we obtain our main result
\be
  v_F(x) = 2 \pi \ell \, \beta(x)\,.
\label{vz2}
\ee
This means that, for the continuum limit, one has effectively $h=\pi \,T$ and all the higher terms in
the series \eqref{h-T} and their expansion coefficients are irrelevant. Since the elements in $T$ vary
parabolically, one recovers in this way the conformal formula for the quantity $\beta(x)$ and also
its prefactor in \eqref{conf_ham}.

The feature that only the first term in \eqref{h-T} contributes, can be discussed in the following way.
In the considerations above, the matrix $T$ has elements depending on position and given
by $z(x)$ in the continuum description. But the form of $z(x)$ does not play a role in the calculation
of $v_F(x)$, so the same result will be obtained for a constant $z(x) \equiv z$. In this case, $T$ is
the hopping matrix of a homogeneous open chain and one can calculate $v_F$ directly as in section\;\ref{sec:cont-limit}.
The eigenvalues of $T$ are $2z\cos(qs)$ and those of $H$ follow as
\begin{equation}
  \nu_q = -N \sum_{m \ge 0} \alpha_m \beta_m \big(2z\cos(qs)\big)^{2m+1} \,.
  \label{nu_hom}
\end{equation}
Looking at $q=\pi/2s$ where $\nu_q$ vanishes, one sees that all terms with $m \ne 0$ have slope zero
there and only $m=0$ remains to give a velocity $v_F=2\pi\ell z$ which is exactly \eqref{vz2}. Thus,
for the homogeneous case, the property $h \simeq \pi \, T$ can be found in a very simple way and is a
direct consequence of the structure of $T$. The calculation above shows that this remains true in a system
which is only locally homogeneous.

\subsection{Finite system}

The same approach can be used for the case of an interval with $N$ sites in a finite ring of
$M$ sites, although there we do not have the analytical form of the entanglement Hamiltonian.
Thus the matrix elements have to be determined numerically following \cite{Peschel03}
and their sums carried out in exactly the same way as it was done in \cite{Arias_etal17_1} for the infinite chain case. 
Aiming directly at $\beta=v_F/2\pi\ell$, we calculate the truncated sum 
\be
  \beta(i) = \frac{1}{\pi N}\, \sum_{p=0}^P (-1)^p (2p+1) H_{i-p,i+p+1}
\label{beta_trunc}  
\ee
where $P$ denotes the cutoff value and the matrix element corresponds to $-t_{2p+1}(i+1/2)$
in the notation of section\;\ref{sec:cont-limit}. This ensures that the 
reflection symmetry on the lattice is respected and that the maxima of the hopping amplitudes
for different values of $p$ are aligned with each other. Due to the finite size of the  subsystem,
one can only obtain and plot $\beta(i)$ in the range $P+1 \le i \le N-(P+1)$.

%
\begin{figure}[bht]
\center
\includegraphics[width=0.49\textwidth]{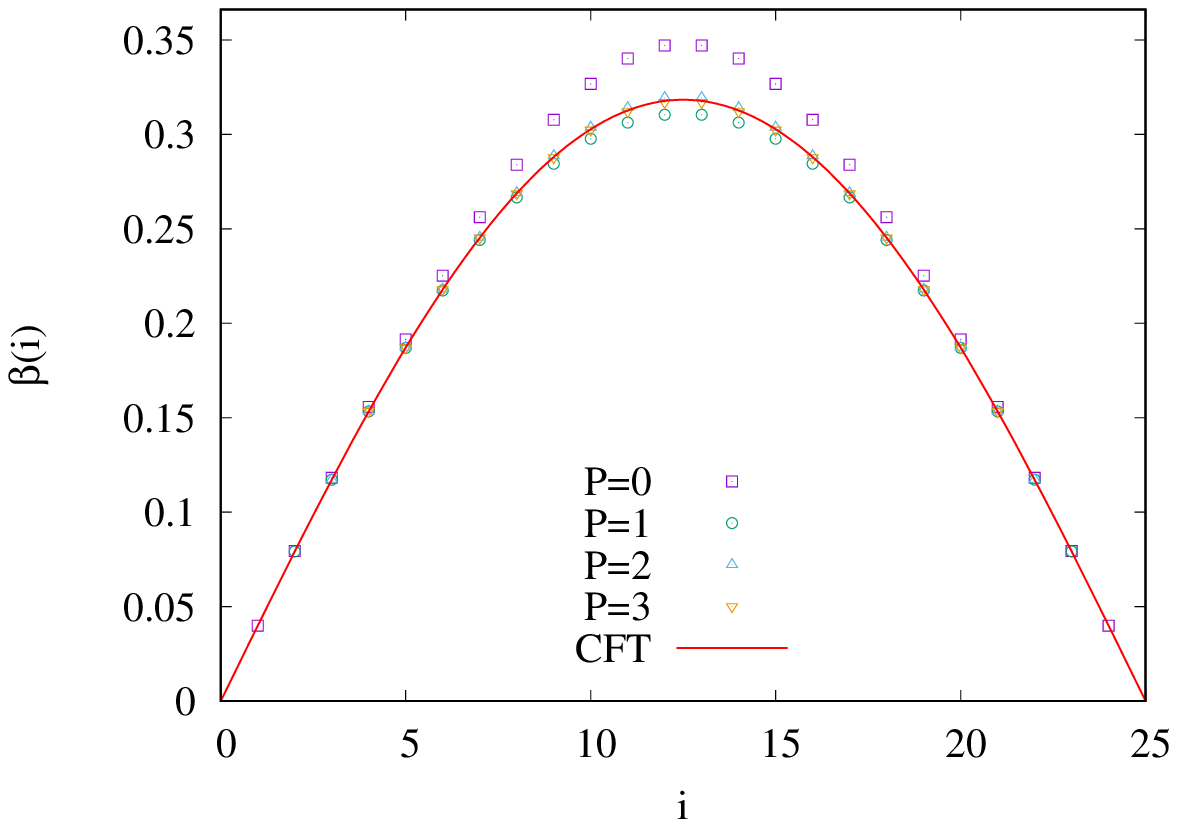}
\includegraphics[width=0.49\textwidth]{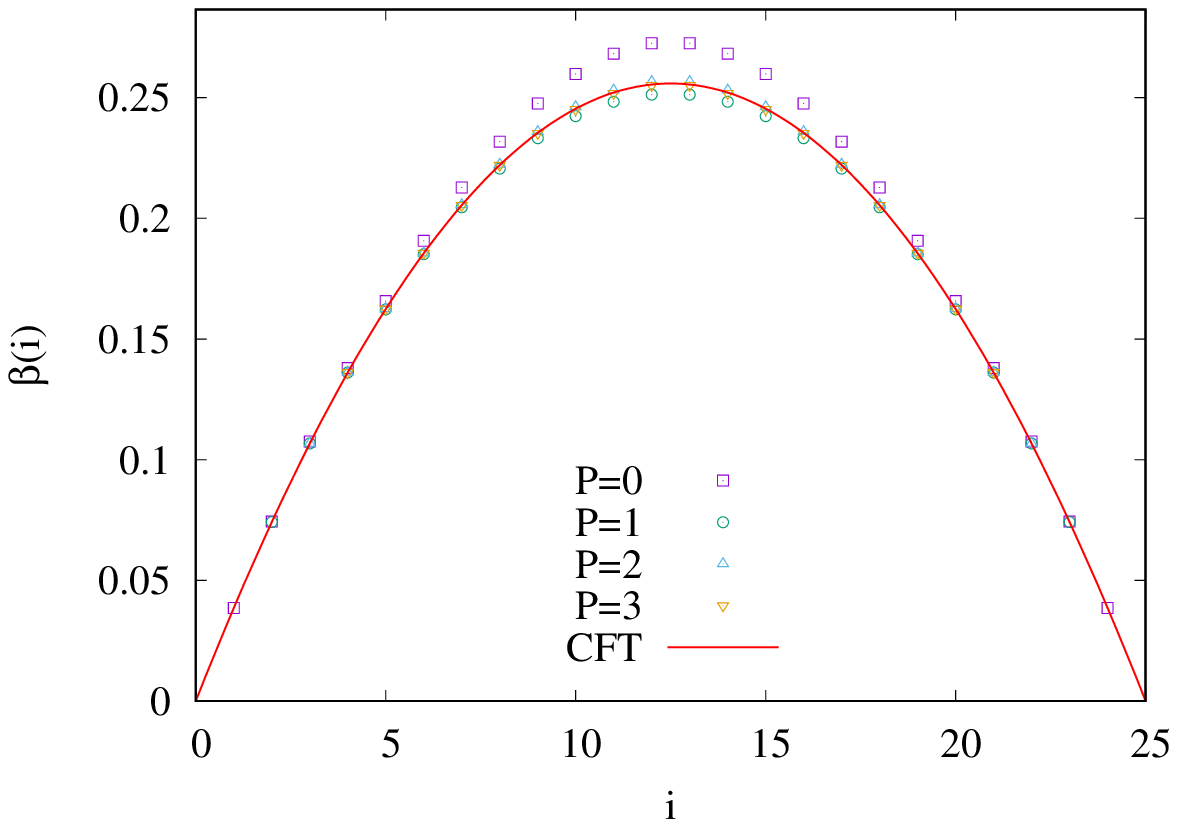}
\caption{The quantity $\beta(i)$ for an interval of $N=25$ sites in a finite ring with two different subsystem
  ratios $r=1/2$ (left) and $r=1/6$ (right). The symbols correspond to different cutoffs $P$ in \eqref{beta_trunc}, while the solid lines show the CFT result \eqref{cftring}. Note the different vertical scales.}
\label{fig:ring}
\end{figure}
%

The result is shown in Fig.\,\ref{fig:ring} for two different ratios $r=N/M=\ell/L$ ($L=Ms$)
and for increasing cutoff values $P$ in the sum. The numerical sums are compared to the CFT result
of \cite{Wong_etal13, Cardy/Tonni16}
\be
\beta(x) = \frac{L}{\pi \,\ell}
            \, \frac{\sin\left(\pi\,x/L\right)\sin\left(\pi(\ell-x)/L\right)}
                  {\sin\left(\pi\,\ell/L\right)}
\label{cftring}
\ee
using $x/L=i/M$ and $\ell/L=N/M$ on the right hand side.
In both cases shown, one can see a clear convergence to the CFT formula. Due to the factor $(-1)^p$ in
the sum, the approximations lie alternately above and below it and already four terms are sufficient
to obtain a very good agreement. For $r=1/6$, one is already relatively close to the case of an infinite
system with the CFT result deviating only slightly from the parabola \eqref{conf_beta}.

\subsection{Finite temperature}

If the chain is infinite but at finite inverse temperature $\beta$, one can proceed in
the same way and calculate $\beta(i)$ via \eqref{beta_trunc}. The corresponding 
CFT formula is obtained from \eqref{cftring} by replacing the sine functions with hyperbolic
sines and the length $L$ of the ring with $\beta$. This gives \cite{Wong_etal13, Cardy/Tonni16}
\be
  \beta(x) =  \frac{\beta}{\pi \,\ell}
              \,\frac{\sinh\left(\pi\,x /\beta\right)\sinh\left(\pi (\ell-x) / \beta\right)}
               {\sinh\left(\pi \,\ell / \beta\right)}
\label{cftbeta}
\ee
with $x=i$ and $\ell=N$ on the right hand side. Note that for small $\beta$, i.e. high temperature,
such that $\ell \gg \beta$, $i \gg \beta$ and $\ell-i \gg \beta$, the quotient of the hyperbolic sines
is approximately $1/2$ and one has $2\pi \ell \beta(i) \simeq \beta $. Thus the local temperature in
the bulk of the subsystem equals the global temperature and its profile shows a pronounced plateau.

\begin{figure}[htb]
\center
\includegraphics[width=0.49\textwidth]{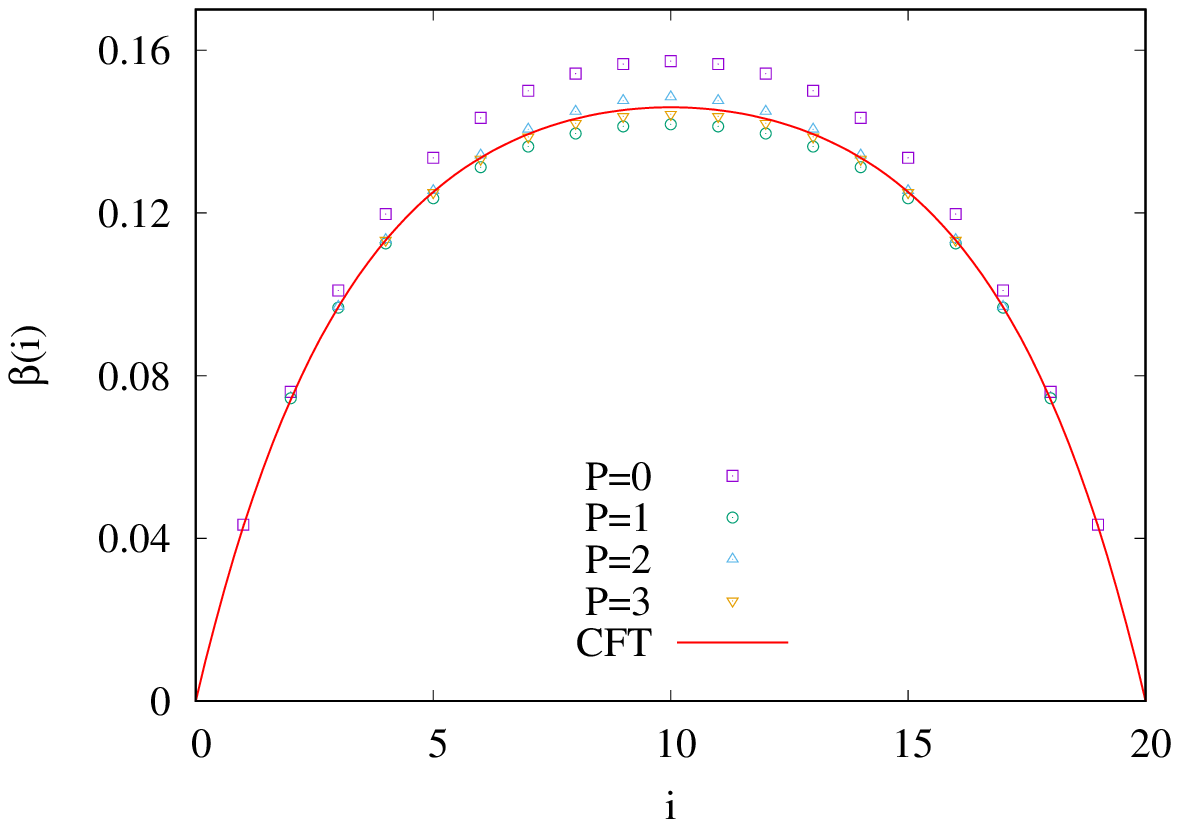}
\includegraphics[width=0.49\textwidth]{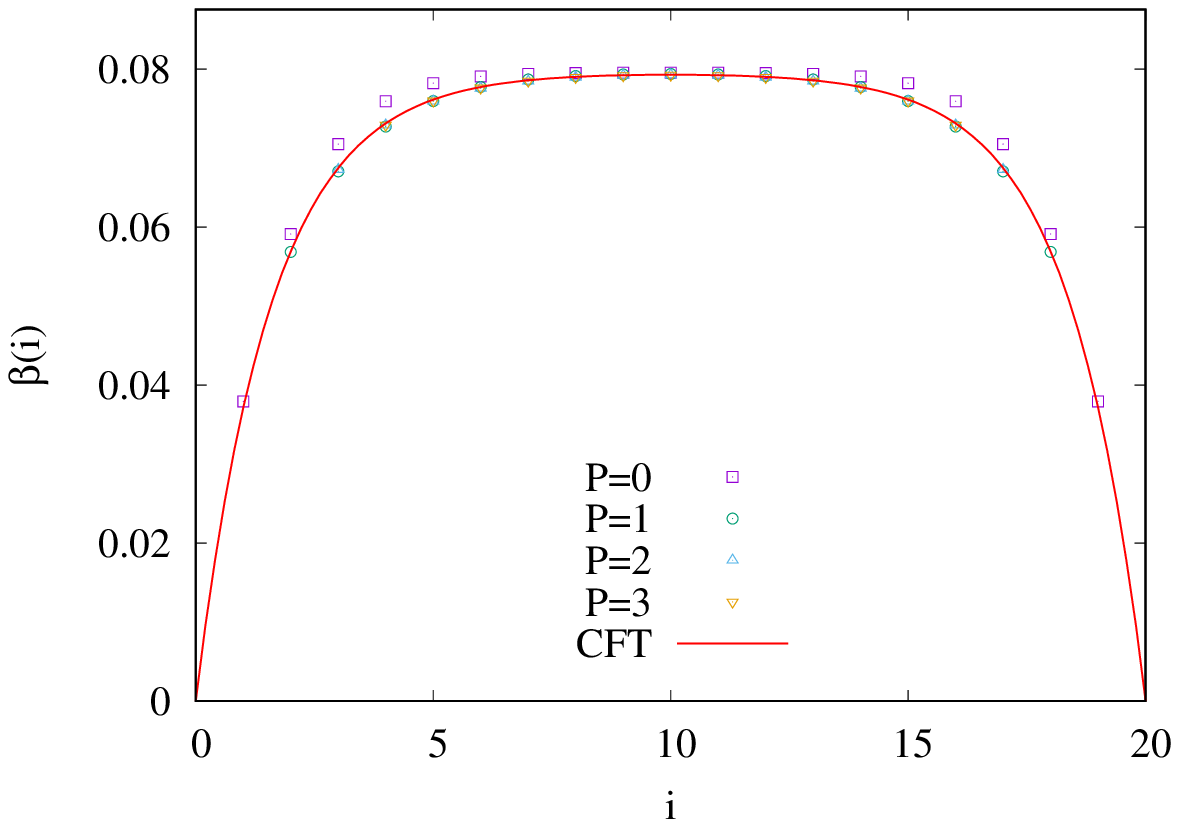}
\caption{The quantity $\beta(i)$ for an interval of $N=20$ sites in an infinite chain at inverse temperature
  $\beta=20$ (left) and $\beta=10$ (right).
  The symbols correspond to different cutoffs $P$ in \eqref{beta_trunc}, while the solid lines show the
  CFT result \eqref{cftbeta}. Note the different vertical scales.}
\label{fig:beta}
\end{figure}

The comparison with the finite sums is shown in Fig.\,\ref{fig:beta},
with again a very good agreement. In particular, one can see that for larger temperatures
the contribution from the higher terms is very small. For $\beta=10$ (right)
it is essentially enough to add the third-neighbour ($p=1$) contribution to recover the CFT result,
while for $\beta=20$ (left) the convergence is slower, similarly to the ground-state case.

%
%

\section{Interval in a chain with arbitrary filling}
\label{sec:arb-filling}

In this section we extend the analysis of section \ref{sec:half-filling} to the case of an infinite chain with arbitrary
filling. This filling will be measured by the Fermi wave number $q_F$ which is related to the average
site occupation via $\bar n = q_F s/\pi $.
The scaled matrix $h=-H/N$ for large $N$ then has the representation \cite{Eisler/Peschel17}
\begin{equation}
  h = \sum_{m \ge 0} \sum_{n=0}^{2m+1}\alpha_m \beta_{m,n}\left(T+\frac{A}{2}\right)^{2m+1-n}
       \left(\frac{A}{2}\right)^n,
       \qquad \quad A=\cos(q_Fs)
  \label{h-Tgen}
\end{equation}
where the matrix $T$ now has also diagonal elements, see Appendix\;\ref{app:T_operator}.
The coefficients $\alpha_m $ and $\beta_{m,n}$ are given in Appendix\;\ref{app:sums} (see (\ref{alpha_beta_mn_def})). 
In the previous notation, the entanglement Hamiltonian then has the form
\be
\label{H_A q_F splitted}
\mathcal{H} =
-\sum_{i} t_0(i) \,c_i^\dagger c_i 
-\sum_{i} \sum_{r\ge 1} t_r(i+r/2)
\Big( 
c_i^\dagger c_{i+r} + c_{i+r}^\dagger c_i 
\Big)
\ee
with on-site terms, hopping to neighbours at arbitrary distances $r$ and no sublattice structure.

To find the continuum limit, we therefore take a different path and introduce right- and left-movers
directly as in bosonization \cite{giamarchi_book}
\bea
\label{c_n continuum q_F}
c_i 
& \longrightarrow &
\sqrt{s} \,\Big(  e^{\textrm{i} q_F x}\, \psi_\textrm{\tiny R}(x) +  e^{-\textrm{i} q_F x}\,  \psi_\textrm{\tiny L}(x) \Big)
\\
\label{c_n shift continuum q_F}
c_{i+r} 
& \longrightarrow &
\sqrt{s} \,\Big(  e^{\textrm{i} q_F (x + r s)}\, \psi_\textrm{\tiny R}(x+ r s) 
+  e^{-\textrm{i} q_F (x + rs)}\,  \psi_\textrm{\tiny L}(x+ r s) \Big)\,.
\eea
Note that this representation with the factors $e^{\pm \textrm{i} q_F x}$ refers to the full chain problem,
although it will be used only in a finite interval.

For the quadratic Hermitean operators occurring in the sums of (\ref{H_A q_F splitted}) one obtains
\bea
\label{c_n term continuum q_F}
c_i^\dagger c_i 
&\;\longrightarrow\;&
s \, \Big( \psi^\dagger_\textrm{\tiny R}(x) \, \psi_\textrm{\tiny R}(x) 
+ \psi^\dagger_\textrm{\tiny L}(x) \, \psi_\textrm{\tiny L}(x)  \Big)
\\
\rule{0pt}{.7cm}
\label{shift term}
c_i^\dagger c_{i+r} + \textrm{h.c.}
&\;\longrightarrow\;&
s \; \Big\{ 
\cos(r q_F s) 
\Big[\,
\psi^\dagger_\textrm{\tiny R}(x) \, \psi_\textrm{\tiny R}(x + r s) 
+  \psi^\dagger_\textrm{\tiny L}(x) \, \psi_\textrm{\tiny L}(x + r s)  
+ \textrm{h.c.}
\,\Big]
\\
&& \hspace{.6cm}
+
\sin(r q_F s) 
\Big[\,
\textrm{i}\,\Big(
\psi^\dagger_\textrm{\tiny R}(x) \, \psi_\textrm{\tiny R}(x + r s) 
-  \psi^\dagger_\textrm{\tiny L}(x) \, \psi_\textrm{\tiny L}(x + r s)  
\Big)
+ \textrm{h.c.}
\,\Big]
\Big\}
\nonumber
\eea
where the terms containing $e^{\pm 2\textrm{i} q_F x}$
have been neglected because they are rapidly oscillating.

To proceed, the $t_r(i+r/2)$ which follow from \eqref{h-Tgen} are necessary.
The  explicit formula is complicated and not essential at this point
(it is given in Appendix\,\ref{app:sums}, see \eqref{trz},\eqref{alpha_beta_mn_def},\eqref{tsr}).
Here we just need to observe that in the continuum limit $t_r(i+r/2)$ becomes a function of $x+r s/2$
(see (\ref{zr})), hence it has to be expanded as $s \to 0$,
finding that $t_r(i+r/2) \,\to\, t_r(x) +s\, r \,t_r'(x) / 2 +O(s^2)$.
Notice that $t_0(i) \longrightarrow t_0(x)$ and that the product $q_F s$ remains constant as $s \to 0$.

By expanding the expressions within the square brackets in (\ref{shift term}) up to $O(s)$
terms included, we obtain for the continuum limit of \eqref{H_A q_F splitted} 
\bea
\label{H_A cont expansion step1}
\mathcal{H}
&\,=&
-\int_0^\ell dx \, t_0(x)\, 
\Big\{ 
\psi^\dagger_\textrm{\tiny R}(x) \, \psi_\textrm{\tiny R}(x) 
+ \psi^\dagger_\textrm{\tiny L}(x) \, \psi_\textrm{\tiny L}(x) 
\Big\}  
-\int_0^\ell dx \, \sum_{r=1}^{\infty} 
\Big(
t_r(x) 
+s\; \frac{r}{2} \,t_r'(x)
\Big)
\\
&&
\times
\bigg\{
\cos(r q_F s) 
\Big[\, 
2\,\big( \psi^\dagger_\textrm{\tiny R}(x) \, \psi_\textrm{\tiny R}(x) + \psi^\dagger_\textrm{\tiny L}(x) \, \psi_\textrm{\tiny L}(x) \big)
+ \, s \,r\;
\partial_x 
\Big(
\psi^\dagger_\textrm{\tiny R}(x) \, \psi_\textrm{\tiny R}(x) + \psi^\dagger_\textrm{\tiny L}(x) \, \psi_\textrm{\tiny L}(x)
\Big)
\,\Big] 
\nonumber
\\
&&
\hspace{0.8cm}
+\sin(r q_F s) 
\Big[\, 
s \,r\,
 \Big(\, \textrm{i}\, \big[\,
\psi^\dagger_\textrm{\tiny R}(x) \, \psi'_\textrm{\tiny R}(x) - \psi^\dagger_\textrm{\tiny L}(x) \, \psi'_\textrm{\tiny L}(x)
\,\big]
+ \,\textrm{h.c.} \,
\Big)
\,\Big] 
\bigg\}\,.
\nonumber
\eea
We remark that an integration by parts leads to a crucial cancellation between the term containing $t_r'(x) $
and the term involving $\partial_x [\, \psi^\dagger_\textrm{\tiny R}(x) \, \psi_\textrm{\tiny R}(x) + \psi^\dagger_\textrm{\tiny L}(x) \, \psi_\textrm{\tiny L}(x) \,]$. 
Thus, (\ref{H_A cont expansion step1}) can be written as
\be
\label{H_A F01}
\mathcal{H}
\,=\,
-\int_0^\ell dx \, F_0(x)\, 
\Big\{ 
\psi^\dagger_\textrm{\tiny R}(x) \, \psi_\textrm{\tiny R}(x) + \psi^\dagger_\textrm{\tiny L}(x) \, \psi_\textrm{\tiny L}(x) 
\Big\}
+
\int_0^\ell dx \,  F_1(x)\,
T_{00}(x)
\ee
where $T_{00}$ is given by (\ref{first T00 def}) and
the functions $F_0(x) $ and $F_1(x) $ are defined as 
\be
\label{F_01 def}
F_0(x) \equiv
t_0(x) + 2 \sum_{r=1}^{\infty}  \cos(r q_F s) \,t_r(x) \,,
\; \qquad \;
F_1(x) \equiv\, 2s  \sum_{r=1}^{\infty}  r \,\sin(r q_F s) \,t_r(x) \,.
\ee
In Appendix\;\ref{app:higherderiv} we perform a systematic analysis of the higher order terms in $s$,
which have been neglected in (\ref{H_A F01}) and involve higher derivatives of the fields.

Compared to the result \eqref{hopham_cont3} in section\;\ref{sec:cont-limit}, the first term in (\ref{H_A F01}) 
containing only densities is new, while in the second one the weight factor in the integral has changed.
Performing these sums by inserting the analytic expressions for $t_r(x) $ is a non trivial task
and the technical details are reported in Appendix\;\ref{app:sums}.
The final result is very simple, however, namely
\be
\label{F_01 sums}
F_0(x) = 0\,,
\;\;\; \qquad \;\;\;
F_1(x) = \, 2\pi \ell \,\beta(x)
\ee
with $\beta(x)$ given by \eqref{conf_beta}. This is completely independent of the filling,
and by inserting \eqref{F_01 sums} into \eqref{H_A F01} one recovers \eqref{hopham_cont3}
with  the expression \eqref{vz2} for $v_F$. In other words, the continuum limit of the entanglement
Hamiltonian has always the same form as for half filling.

This result can be understood as in the previous section by considering a system which is
homogeneous. The matrix
$T$ is now a hopping matrix with diagonal terms. If its elements are constant, it has eigenvalues
$2z(\cos(qs)-A)$ and those of $H$ follow as
\begin{equation}
  \nu_q = -\,N \sum_{m \ge 0} \sum_{n=0}^{2m+1}\alpha_m \beta_{m,n}
          \left[ \, 2z(\cos(qs)-A)+\frac{A}{2}\,\right]^{2m+1-n}\left(\frac{A}{2}\right)^n.
  \label{nu_hom_gen}
\end{equation}
This expression can be shown to vanish for $q=q_F$, and the velocity at this point is given by
\begin{equation}
  v_F = 2 \ell z \sin(q_Fs) \sum_{m \ge  0} \sum_{n=0}^{2m+1}\alpha_m \beta_{m,n} (2m+1-n)
          \left(\frac{A}{2}\right)^{2m}.
  \label{v_hom_gen}
\end{equation}
Inserting the values of the coefficients given in Appendix\;\ref{app:sums}, one finds that the sums can be
carried out as 
\begin{equation}
  \sum_{n=0}^{2m+1} \beta_{m,n} (2m+1-n) \left(\frac{1}{2}\right)^{2m}=\pi\,, 
  \qquad
  \sum_{m \ge 0}\alpha_m A^{2m}=\frac{1}{\sqrt{1-A^2}}=\frac{1}{\sin(q_Fs)}
  \label{sums_gen}
\end{equation}
and the final result is $v_F = 2 \pi \ell z$ as for half filling. The mechanism is somewhat different, however. 
While for $A=0$ only the term $m=0$ remains, one needs here the full sum over $m$, and this cancels the quantity
$\sin(q_Fs)$ which appears initially.


\section{Conclusions}
\label{sec:conclusions}

We have studied the question, how the results for the entanglement Hamiltonian in discrete
hopping chains can be reconciled with the predictions of conformal field theory. For this, we
first found out how the long-range couplings of the discrete system enter into the continuum limit
and then evaluated the corresponding sums. This was quite transparent for a half-filled infinite
chain, but considerably more involved for general filling. In both cases, the conformal expression
could be reobtained analytically.

The main ingredient was a formula which expresses the hopping matrix $H$ in $\mathcal{H}$ as a power
series of a commuting, in general tridiagonal matrix $T$. For half filling, the  mechanism found 
was that while the individual hopping amplitudes in $H$ have contributions from various powers
of $T$, almost all of these cancel in the superposition which is needed for the continuum limit and
the local velocity $v_F(x)$ is completely determined by the first power. The desired continuum
result is then obtained, because $T$ has the conformal form already on the lattice. Turning the
argument around, one could say that it is the continuum limit which demands this particular
property of $T$.

One should mention that the relation $H \simeq -\,\pi N T$ has been encountered before in a related
problem, namely for the low-lying eigenvalues of the two quantities \cite{Eisler/Peschel17,Peschel04}.
Assuming that these are also the relevant ones for the continuum limit, this is an anticipation
of the result found here for the Hamiltonian itself. For general filling, the relation is changed
to $H \simeq - \,\pi N T/\sin(q_Fs)$, where the sine factor can be viewed as the Fermi velocity in the
physical Hamiltonian.
Indeed, as shown in Appendix\;\ref{app:T_operator}, the above relation becomes exact between the
operators $\mathcal{H}$ and $\mathcal{T}$, if one considers both of them in the continuum limit.

We have considered also the case of a finite ring and found that the approach works equally well
there and the conformal result is recovered numerically. In this case, no explicit formula like
\eqref{h-T} has been derived, but a commuting matrix $T$ is known and its elements have the conformal
form. So the mechanism seems to be the same in this case. This is in line with the results for the
low-lying eigenvalues. The same is expected to hold for open chains, where again a commuting $T$
exists \cite{Eisler/Peschel18}.

Taking all this together, our calculations both show how the discrete and the continuum results
are connected in these free-fermion systems and how the particular commuting operator which exists
here enters into the considerations. We think that this sheds additional light onto this algebraic
structure.

We have limited ourselves here to chains which were homogeneous, but there are also conformal
predictions for the entanglement Hamiltonian in inhomogeneous chains \cite{Tonni/Laguna/Sierra17}.
It would be interesting to see, how it can be obtained from the lattice result in such a situation,
for example in the so called rainbow chain \cite{Vitagliano/etat10, Ramirez/Laguna/Sierra14},
where the hopping decreases exponentially from the centre of the system. 
Whether a simple commuting operator exists in this case, is not known.

Finally, it would be interesting to consider also interacting lattice models, where analytical
results on the entanglement Hamiltonians are still missing, and a direct numerical evaluation
is limited to very small system sizes \cite{Nienhuis/Campostrini/Calabrese09}.
To overcome this barrier, various numerical approaches have been proposed recently
\cite{Parisen/Assaad18,Zhu/Huang/He18,Dalmonte/Vermersch/Zoller17,Giudici/etal18}.
The results for the low-lying eigenvalues suggest that the continuum limit derived here for
free fermions should also apply to more complicated systems.

\section*{Acknowledgements}

\noindent
We would like to thank Ra\'ul Arias and Horacio Casini for inspiring discussions.
The work reported here was started during the programme ``Entanglement in Quantum Systems'' in Florence
in June and July 2018 and VE and IP thank the Galileo Galilei Institute for the invitation and the hospitality.
VE acknowledges funding from the Austrian Science Fund (FWF) through Project No. P30616-N36.



\begin{appendices}

\section*{Appendices}

\section{Continuum limit of the operator $\mathcal{T}$}
\label{app:T_operator}

In this appendix we consider the continuum limit of the operator $\mathcal{T}$
\begin{equation}
  \mathcal{T}=  \sum_{i,j=1}^N \, T_{i,j} c^{\dag}_i c_j \, 
\label{T_op_def}
\end{equation}
formed with the matrix $T$ which occurred repeatedly in the main text.
Because $T$ commutes with $H$ in (\ref{ent_ham}), the operator $\mathcal{T}$ commutes with $\mathcal{H}$.
Given the matrix elements of $T$ and in the previous notation, for arbitrary
filling  it reads
\be
\label{T_A operator}
\mathcal{T} =
\sum_{i=1}^N d(i) \,c^\dagger_i\, c_i
+
\sum_{i=1}^{N-1} t(i) \big( c^\dagger_i\, c_{i+1} + c^\dagger_{i+1}\, c_i \big) 
\ee
where
\be
\label{d_n and t_n def}
d(i) \equiv\, - \,2\cos(q_Fs)\, \frac{2i - 1}{2N} \left( 1 - \frac{2i - 1}{2N} \right)\,,
\;\; \qquad \;\;
t(i) \equiv  \frac{i}{N} \left( 1 -  \frac{i}{N} \right).
\ee
The continuum limit of $\mathcal{T}$ can be studied by following the same steps discussed
for the entanglement Hamiltonian $\mathcal{H} $ in section\;\ref{sec:arb-filling}. However, 
the analysis is simpler because the hopping term involves only the nearest neighbours
and it is useful to present it separately.

If one uses $ (i-1/2)s \to x$ in the continuum limit, one finds 
$ d(i)  \to -\, 2\cos(q_F s)\,\beta(x)$, with $\beta(x)$ given by (\ref{conf_beta}). 
The continuum limit of the coefficient $t(i)$ is also straightforward, once it
is written in the following form
\bea
t(i) 
&=&
 \frac{i-1/2}{N} \left( 1 -  \frac{i-1/2}{N}  \right)
 + \frac{1/2}{N} \left( 1 -  2\; \frac{i-1/2}{N}  \right)
 - \left(\frac{1/2}{N}  \right)^2
 \\
 && 
   \longrightarrow \;\;
\beta(x)
+ \frac{\beta'(x)}{2} \, s\,
-  \left( \frac{1/2}{\ell} \right)^2 s^2.
\nonumber
\eea
The continuum limit of the operators in (\ref{T_A operator}) is obtained by using the above observations,
by introducing the fields $\psi_\textrm{\tiny L}(x) $ and $\psi_\textrm{\tiny R}(x) $ 
as in  (\ref{c_n continuum q_F}), and by employing (\ref{c_n shift continuum q_F}) with $r=1$.
The result reads
\bea
\label{T operator continuum v1}
N\, \mathcal{T}
& \longrightarrow &
-\, 2\,N \sum_{n=1}^N s\, \cos(q_Fs)\,\beta(x)\,
\Big\{ 
\psi^\dagger_\textrm{\tiny R}(x) \, \psi_\textrm{\tiny R}(x) + \psi^\dagger_\textrm{\tiny L}(x) \, \psi_\textrm{\tiny L}(x) 
\Big\}
\\
\rule{0pt}{.8cm}
&&
+ \,N \sum_{n=1}^N s\,
\bigg( 
\beta(x) + \frac{\beta'(x)}{2} \; s\, 
\bigg) 
\nonumber 
\\
&& 
\times \bigg\{\!
\cos(q_Fs) \, 
\Big[\,
2\,\big( \psi^\dagger_\textrm{\tiny R}(x) \, \psi_\textrm{\tiny R}(x) + \psi^\dagger_\textrm{\tiny L}(x) \, \psi_\textrm{\tiny L}(x) \big)
+ \, s \;
\partial_x 
\Big(
\psi^\dagger_\textrm{\tiny R}(x) \, \psi_\textrm{\tiny R}(x) + \psi^\dagger_\textrm{\tiny L}(x) \, \psi_\textrm{\tiny L}(x)
\Big)
\,\Big]
\nonumber 
\\
&& \hspace{5cm}
 +\, s\,  \sin(q_F) \; 
\Big[\,
\textrm{i}\, \big(\,
\psi^\dagger_\textrm{\tiny R}(x) \, \psi_\textrm{\tiny R}(x)' - \psi^\dagger_\textrm{\tiny L}(x) \, \psi_\textrm{\tiny L}(x)'
\,\big)
+ \,\textrm{h.c.} 
\,\Big]
\bigg\}\,.
\nonumber 
\eea
It is straightforward to notice that the two terms at leading order cancel;
hence we are left with
\bea
\label{T operator continuum v1 next}
N\, \mathcal{T}
& \longrightarrow \;\;&
\ell  \int_0^\ell  \bigg\{\,
\beta'(x)\, \cos(q_F s) \, \big( \psi^\dagger_\textrm{\tiny R}(x) \, \psi_\textrm{\tiny R}(x) + \psi^\dagger_\textrm{\tiny L}(x) \, \psi_\textrm{\tiny L}(x) \big)
\\
& &\hspace{1.2cm}
+  \,\beta(x)\,\cos(q_F s) \, 
\partial_x 
\Big(
\psi^\dagger_\textrm{\tiny R}(x) \, \psi_\textrm{\tiny R}(x) + \psi^\dagger_\textrm{\tiny L}(x) \, \psi_\textrm{\tiny L}(x)
\Big)
\nonumber 
\\
& &\hspace{1.2cm}
+ \,\beta(x) \, \sin(q_F s) \; 
\Big[\,
\textrm{i}\, \big(\,
\psi^\dagger_\textrm{\tiny R}(x) \, \psi'_\textrm{\tiny R}(x) - \psi^\dagger_\textrm{\tiny L}(x) \, \psi'_\textrm{\tiny L}(x)
\,\big)
+ \,\textrm{h.c.} 
\,\Big]
\bigg\} \, dx\,.
\nonumber 
\eea
The first two lines in this expression cancel once an integration by parts is performed.
Thus, the final result reads
\be
\label{T_A continuum limit final}
N\, \mathcal{T}
\;\;\longrightarrow \;\;
\sin(q_F s) \; \ell\int_0^\ell dx\,
\beta(x)\, 
\Big[\,
\textrm{i}\, \big(\,
\psi^\dagger_\textrm{\tiny R}(x) \, \psi'_\textrm{\tiny R}(x) - \psi^\dagger_\textrm{\tiny L}(x) \, \psi'_\textrm{\tiny L}(x)
\,\big)
+ \,\textrm{h.c.} 
\,\Big]\, .
\ee
This shows that the rescaled operator $ -\,\pi N\,\mathcal{T} / \sin(q_F s)$
has the same continuum limit as the entanglement Hamiltonian $\mathcal{H}$. Such a relation also
exists for the low-lying eigenvalues of the corresponding matrices $T$ and $H$ \cite{Eisler/Peschel17}.

Finally, let us observe that, by defining $x$ in the continuum limit 
as $ is \to x$ instead of $ (i-1/2)s \to x$, one finds
\be
d(i) \; \longrightarrow \; -\, 2\cos(q_F s)
\left[\,
\beta(x)
- \frac{\beta'(x)}{2} \; s
-  \left( \frac{1/2}{\ell} \right)^2 s^2
\,\right],
\;\; \qquad \;\;
t(i) \; \longrightarrow \; \beta(x)\,.
\ee 
Then, after some simplifications similar to the ones discussed above, 
the limit (\ref{T_A continuum limit final}) is recovered, as expected.

\section{Calculation of the sums for arbitrary filling}
\label{app:sums}

In this appendix we prove the identities (\ref{F_01 sums}) for the sums
defined in (\ref{F_01 def}).

As noted in the main text, the $r$-th neighbour hopping amplitude $t_r(i)= N h_{i,i+r}$
in the entanglement Hamiltonian is actually a function of the scaling variable
\be
z_{r} = \frac{2i+r-1}{2N}\left(1-\frac{2i+r-1}{2N}\right) .
\label{zr}
\ee
The explicit analytic expression in this scaling limit was found in \cite{Eisler/Peschel17} and reads
\be
\frac{1}{N} \, t_r(z_r) = \sum_{m=0}^{\infty} \sum_{n=0}^{2m+1}
\alpha_m \beta_{m,n}
\left(\frac{A}{2}\right)^n
\tilde t_{2m+1-n,r}(z_r)
\label{trz}
\ee
where the coefficients $\alpha_m$ and $\beta_{m,n}$ are given by
\be
\alpha_m = \frac{1}{\sqrt{\pi}} \frac{\Gamma(m+1/2)}{\Gamma(m+1)}\;, 
\qquad
\beta_{m,n} = 2^{2m} \frac{\Gamma(2m-n+1/2)\Gamma(n+1/2)}{\Gamma(2m-n+2)\Gamma(n+1)} 
\label{alpha_beta_mn_def}
\ee
whereas $A=\cos (q_F s)$ and
\be
\tilde t_{k,r}(z_r) = \sum_{\substack{\ell=0  \\ k-r-\ell \,\, \mathrm{even}}}^{k-r}
\binom{k-\ell}{\frac{k-r-\ell}{2}} \binom{k}{\ell} \left(\frac{A}{2}-2Az_r \right)^\ell z_r^{k-\ell}\,.
\label{tsr}
\ee
The appearance of the scaling variable $z_r$ follows from simple symmetry reasons,
by requiring that $h_{i,i+r}$ be invariant under the reflection $i+r \to N+1-i$.
However, as explained in the main text, when carrying out the continuum limit it is
easier to take these shifts into account via the expansion \eqref{H_A cont expansion step1},
and work instead with the scaling variable $z \equiv z_0$ for
each $r$. Using $N s = \ell$ and $z=\beta(x)$, the identities \eqref{F_01 sums} we have to prove are
\be
t_0(z) + 2\sum_{r=1}^{\infty} \cos(r q_F s) \, t_{r}(z)=0\,, 
\qquad
\sum_{r=1}^{\infty} r \sin(r q_F s) \, t_r(z) = N \pi z \,.
\label{tsum}
\ee

The strategy one should follow is essentially the same as for half filling, but the calculation
is much more cumbersome. First one rewrites the hoppings as a power series
\be
t_{r}(z) = N \sum_{p=r}^{\infty} \gamma_{p,r} z^p \, .
\label{tzser}
\ee
In fact, the lowest order is always given by the range $r$ of the hopping, which becomes
clear by rewriting the sum \eqref{tsr} after the substitution $\ell \to k-\ell$ as
\be
\tilde t_{k,r}(z) = \sum_{\ell=r}^{k}{}'
\binom{\ell}{\frac{\ell-r}{2}} \binom{k}{\ell} \left(\frac{A}{2}-2Az \right)^{k-\ell} z^{\ell}\,.
\label{tsr2}
\ee
The prime over the sum denotes that the summation index $\ell$ must have the same parity as $r$.
To extract the coefficient $\gamma_{p,r}$ of $z^p$, one should expand
\be
\left(\frac{A}{2}-2Az \right)^{k-\ell} 
=\; 
\sum_{j=0}^{k-\ell}
\binom{k-\ell}{j} \left(\frac{A}{2}\right)^{j}(-2Az)^{k-\ell-j}\,.
\ee
Clearly, in order to produce the power $z^p$, one has to match the factor $z^\ell$
with a term of order $z^{p-\ell}$, such that we need only the term satisfying $k-j=p$.
The prefactor of $z^p$ in $\tilde t_{k,r}(z)$ is then 
\be
\sum_{\ell=r}^{p}{}' \binom{\ell}{\frac{\ell-r}{2}} \binom{k}{\ell}
\binom{k-\ell}{k-p} \left(\frac{A}{2}\right)^{k-p} (-2A)^{p-\ell}\,.
\label{tsrzp}
\ee
Note that we need to have $k \ge p$.
Since the $z$ dependence of $t_r(z)$ is entirely due to the terms
$\tilde t_{2m+1-n,r}(z)$, one can now set $k=2m+1-n$ and plug the factors 
\eqref{tsrzp} into the sum \eqref{trz} to get the prefactor $\gamma_{p,r}$. 
This yields
\be
\gamma_{p,r} = \left[ \sum_{m=0}^{\infty} \sum_{n=0}^{2m+1-p}
\alpha_m \beta_{m,n} \left(\frac{A}{2}\right)^{2m+1-p}
\frac{(2m+1-n)!}{(2m+1-n-p)!} \right] \, C_{p,r}\, = \sigma_p \, C_{p,r}
\label{gpr}
\ee
where the quantity $\sigma_p$ in the brackets only depends on $p$ and we have defined 
\be
C_{p,r}= \sum_{\ell=r}^{p}{}' \frac{(-2A)^{p-\ell}}
{\left(\frac{\ell-r}{2}\right)!\left(\frac{\ell+r}{2}\right)!(p-\ell)!}=
\sum_{k=0}^{\lfloor \frac{p-r}{2} \rfloor} \frac{(-1)^{p-r}(2A)^{p-r-2k}}
{k!(k+r)!(p-r-2k)!}\,.
\label{cpr}
\ee
The second equality above follows by substituting $k=(\ell-r)/2$. Two special values of $\sigma_p$
are
\be
  \sigma_0 = 0\,,\quad \quad \sigma_1 = \frac{\pi}{\sin (q_F s)}
  \label{sigma}
  \ee
where the first one follows from the relation $\sum_{n=1}^{2m+1}\beta_{m,n}=0$ and the second one
from the equations \eqref{sums_gen}.

Having the power series \eqref{tzser} at hand, the identities \eqref{tsum} now translate into
\be
\gamma_{p,0} + 2 \sum_{r=1}^{p} \cos(r q_F s) \gamma_{p,r}=0\,, 
\qquad
\sum_{r=1}^{p} r \sin(r q_F s) \, \gamma_{p,r} =  \pi \, \delta_{p,1} \, .
\label{gsum}
\ee
In other words, we have to prove that the prefactor of each power $z^p$ in \eqref{tsum} vanishes, except for
the linear term $p=1$ in the second sum. For $p=0$, the sums are absent since then $p<r$. In this case,
the left equation holds because $\sigma_0=0$ and the right one is trivial. Using \eqref{gpr}, the relations
\eqref{gsum} can be transformed into the following ones for the $C_{p,r}$
\be
C_{p,0} + 2 \sum_{r=1}^{p} \cos(r q_F s) C_{p,r}=\delta_{p,0} \, , 
\qquad
\sum_{r=1}^{p} \frac{\sin(r q_F s)}{\sin (q_F s)} \, r \, C_{p,r} =  \delta_{p,1} \, ,
\label{csum}
\ee
where in the first equation $\delta_{p,0}$ is due to the fact that $C_{0,0}=1$ 
and in the second one $\sigma_1$ from \eqref{sigma} has been used.

In order to prove \eqref{csum}, one should note that each factor $C_{p,r}(A)$
in \eqref{cpr} is given as a polynomial in terms of the parameter $A$.
Thus, we will need a similar representation of the trigonometric factors as well.
This is accomplished by recognizing them as the Chebyshev polynomials
$T_r$ and $U_r$  \cite{Gradshteyn/Ryzhik65}, which for $r\ne 0$ can be written as
\be
\cos (r q_F s) = T_r(A) = \frac{r}{2} \sum_{k=0}^{\lfloor \frac{r}{2} \rfloor}
(-1)^k \frac{(r-1-k)!}{k!(r-2k)!} (2A)^{r-2k}
\label{cqr}
\ee
and
\be
\frac{\sin (r q_F s)}{\sin (q_F s)} = U_{r-1}(A) = 
\sum_{k=0}^{\lfloor \frac{r-1}{2} \rfloor}
(-1)^k \frac{(r-1-k)!}{k!(r-1-2k)!} (2A)^{r-1-2k}\,.
\label{sqr}
\ee
Dropping now the arguments $A$, we have to prove
\be
C_{p,0} + 2 \sum_{r=1}^{p} T_r \, C_{p,r}=\delta_{p,0} \, , \qquad
\sum_{r=1}^{p} r \, U_{r-1} \, C_{p,r} =  \delta_{p,1} \, .
\label{csum2}
\ee

First we show that the second identity follows simply from the first one
by taking the derivative with respect to $A$. Indeed, one has
\be
C'_{p,0} + 2 \sum_{r=1}^{p-1} T_r \, C'_{p,r} + 2 \sum_{r=1}^{p} T'_r \, C_{p,r} = 0 \, , \qquad
\label{dcsum}
\ee
Furthermore one has
\be
T'_r = r \, U_{r-1}\,, 
\qquad
C'_{p,r} = -2 C_{p-1,r}
\ee
where the first one is a well-known identity between Chebyshev polynomials and the
second one can be easily verified using the definition \eqref{cpr}.
Substituting into \eqref{dcsum} and using the first (yet unproven) identity
from \eqref{csum2} one gets
\be
2 \sum_{r=1}^{p} r \, U_{r-1} \, C_{p,r} =
2(C_{p-1,0} + 2 \sum_{r=1}^{p-1} T_r \, C_{p-1,r})=
 2 \delta_{p-1,0} = 2\delta_{p,1}\,.
\ee

Thus it remains to prove the first identity of \eqref{csum2}, which essentially
amounts to collect the various powers of $A$ in the product of polynomials.
Indeed, one should note that the $C_{p,r}$ are polynomials of order $p-r$,
where only terms corresponding to that parity appear. Similarly, the Chebyshev
polynomials $T_r$ are of the order $r$, containing only terms with that parity.
Thus, for each $r$, the highest power that appears in their product is $p$,
corresponding to the $k=0$ term in the sums \eqref{cpr} and \eqref{cqr}.
Collecting also the contribution from $C_{p,0}$ and summing over $r$,
the $p$-th order term in the polynomial is given by
\be
\left[\frac{1}{p!} + \sum_{r=1}^p \frac{(-1)^r}{r!(p-r)!} \right] (-1)^p (2A)^p
= \delta_{p,0} \, .
\ee
Hence, we see that the highest order terms already deliver the desired result.

The last step is to prove that each of the remaining powers sum up to zero.
Let us consider the contribution of the $(p-2\ell$)-th power for a fixed $\ell > 0$.
This can combine as $(p-r-2k) + (r-2(\ell-k))$, i.e. each term $k$ in the sum of $C_{p,r}$ must
combine with $k' = \ell-k$ in that of $T_r$. The corresponding contribution of their product then reads
\be
\frac{(-1)^{\ell-k+r} r \, (r-1-(\ell-k))!}
{k!(k+r)!(p-r-2k)!(\ell-k)!(r-2(\ell-k))!} 
(-1)^p (2A)^{p-2\ell}\,.
\ee
When collecting the contributions from the various $k$ and $r$, 
one has to take care of the limits. Here we would like to treat $k$ as an
independent summation variable. To ensure that $k' \ge 0$, one has to choose
$0 \le k \le \ell$. Note, however, that the original range of hopping $1 \le r \le p$ must
also be restricted. Indeed, one has to ensure that both powers in the product
be positive, $p-r-2k \ge 0$ and $r-2(\ell-k) \ge 0$, thus the range of summation
over $r$ depends on the value of $k$. Collecting also the contribution of the
term $C_{p,0}$, the overall prefactor of the term $(-1)^p (2A)^{p-2\ell}$ becomes
\be
\frac{1}{(\ell!)^2(p-2\ell)!} + \sum_{k=0}^{\ell}
\sum_{r=\max \left(2(\ell-k),1\right)}^{p-2k}
\frac{(-1)^{\ell-k+r} r \, (r-1-(\ell-k))!}
{k!(k+r)!(p-r-2k)!(\ell-k)!(r-2(\ell-k))!} \, .
\label{Apf}
\ee

One can see that in the above sum over $r$ the $k=\ell$ term requires special attention,
since for this value the lower limit $2(\ell-k)$ would give zero, instead of the requirement $r \ge 1$.
Thus, we first consider $k<\ell$ and introduce the new variables $r'=r-2(\ell-k)$ and $k'=\ell-k$, such that
the sum can be rewritten as
\be
\sum_{k'=1}^{\ell} \sum_{r'=0}^{p-2\ell}
\frac{(-1)^{k'+r'} (r'+2k') \, (r'+k'-1)!}
{(l-k')!(r'+k'+\ell)!(p-r'-2\ell)! \, k'! \, r'!} \,.
\label{sum1}
\ee
Now, since the summation ranges are independent of each other, we can
interchange them. It turns out that the $k'$-sum can then be carried out as
\be
\sum_{k'=1}^{\ell}
\frac{(-1)^{k'} (r'+2k') \, (r'+k'-1)!}{k'!(\ell-k')!(r'+k'+\ell)!} =
-\frac{r'!}{\ell!(\ell+r')!}\,.
\label{sum2}
\ee
Including now also the $k=\ell$ term which was left out, the sums in \eqref{Apf}
can be rewritten as
\be
\sum_{r=1}^{p-2\ell} \frac{(-1)^{r}} {\ell!(\ell+r)!(p-r-2\ell)!} -
\sum_{r'=0}^{p-2\ell} \frac{(-1)^{r'}} {\ell!(\ell+r')!(p-r'-2\ell)!} =
-\frac{1}{(\ell!)^2(p-2\ell)!}\,.
\label{sum3}
\ee
Since this is exactly the opposite of the first term in \eqref{Apf}, the prefactor
vanishes for arbitrary $p$ and $\ell>0$. This concludes our proof of \eqref{csum2}.


\section{Higher derivatives}
\label{app:higherderiv}

The result (\ref{H_A F01}) for the continuum limit of (\ref{H_A q_F splitted}) involves only the 
zero-th and first order derivatives of the fields. 
In this appendix we consider also the subleading terms, which involve higher order derivatives of the fields
and have been neglected in (\ref{H_A cont expansion step1}) because they vanish in the continuum limit. They
are interesting, however, since they lead to closely related expressions.

As for the coefficients in (\ref{H_A q_F splitted}), 
we have that $t_0(i) \longrightarrow t_0(x)$, while $t_r(i+r/2) $ provides the following derivative expansion in the continuum limit
\be
\label{expansion t_r all}
t_r(i+r/2) 
\;\, \longrightarrow \; \,
t_r(x) + \sum_{l = 1}^{\infty}
\frac{(s\,r/2)^l}{l!}\; t_r^{(l)}(x)
\qquad
r\ge 1
\ee
Furthermore, let us consider the expansion of the expression within the curly brackets in (\ref{shift term}), 
which can be written as
\be
\label{expansion psi-psi all} 
-2\cos(r q_F s) 
\sum_{k = 0}^{\infty}
\frac{(r s)^k}{k!}\,\Psi^{(k)}_+ 
-
2\sin(r q_F s) 
\sum_{k = 0}^{\infty}
\frac{(r s)^k}{k!}\,\Psi^{(k)}_- 
\ee
where we have introduced the following quadratic expressions of the fields 
\be
\label{Psi's def}
\Psi^{(k)}_+ 
\equiv\,
-\frac{1}{2}\,
\Big(
\psi^\dagger_\textrm{\tiny R} \, \psi_\textrm{\tiny R}^{(k)}
+ \psi^\dagger_\textrm{\tiny L} \, \psi_\textrm{\tiny L}^{(k)}
\Big)
+ \textrm{h.c.}\,,
\,\qquad\,
\Psi^{(k)}_- 
\equiv\,
-\frac{\textrm{i}}{2}\,\Big(
\psi^\dagger_\textrm{\tiny R} \, \psi_\textrm{\tiny R}^{(k)}
- \psi^\dagger_\textrm{\tiny L} \, \psi_\textrm{\tiny L}^{(k)}
\Big)
+ \textrm{h.c.}
\ee
being $k  \ge 0$ and $\psi^{(k)}\equiv\partial_x^{(k)}\psi(x)$.
Notice that for $k=0$ we have $\Psi^{(0)}_+ = -(\psi^\dagger_\textrm{\tiny R} \, \psi_\textrm{\tiny R}+ \psi^\dagger_\textrm{\tiny L} \, \psi_\textrm{\tiny L}) $ 
and $\Psi^{(0)}_- = 0$.
Instead, for $k=1$ we recognise $\Psi^{(1)}_- = T_{00}$ defined in (\ref{first T00 def}).

The continuum limit of the entanglement Hamiltonian (\ref{H_A q_F splitted}) can be studied by employing 
(\ref{c_n term continuum q_F}), (\ref{expansion t_r all}) and (\ref{expansion psi-psi all}), finding that
\bea
\label{H_A cont expansion all derivatives}
\mathcal{H} 
&  =& 
\int_0^\ell dx\, t_0\, \Psi^{(0)}_+
\\
\rule{0pt}{.8cm}
& &
+\,2 \sum_{r=1}^{\infty} \int_0^\ell dx \,
\Bigg(
\sum_{l = 0}^{\infty}
\frac{(s\,r)^l}{q!\, 2^l }\; t_r^{(l)}
\Bigg) \,
\sum_{k = 0}^{\infty}
\frac{(r s)^k}{k!}\,
\Big[
\cos(r q_F s)  \Psi^{(k)}_+ 
+
\sin(r q_F s) \,\Psi^{(k)}_- 
\,\Big]
\nonumber
\eea
which reduces to (\ref{H_A cont expansion step1}) when  the $o(s)$ terms are neglected. 
The $r$-th term of the sum in the r.h.s. in (\ref{H_A cont expansion all derivatives}) can be written as follows
\be
\label{inter-term}
 2 \int_0^\ell dx\,
 \sum_{m = 0}^{\infty} \frac{(r s)^m}{m!\, 2^m }\;
\sum_{k = 0}^{m} 2^k  \binom{m}{k} \,t_r^{(m-k)} 
\Big[
\cos(r q_F s)  \Psi^{(k)}_+ 
+
\sin(r q_F s) \,\Psi^{(k)}_- 
\,\Big]
.
\ee
In each term of this expression, let us perform the proper number of integrations by parts in order to isolate the functions $t_r$
and plug the result back into (\ref{H_A cont expansion all derivatives}).
Then, by taking into account also the sum in $r$ in (\ref{H_A cont expansion all derivatives})
and exchanging the order of the two sums in the second integral, 
we find that the entanglement Hamiltonian (\ref{H_A cont expansion all derivatives}) can be written as follows
\be 
\label{H high der m-sum}
\mathcal{H} 
\,=\, 
\sum_{m = 0}^{\infty} \frac{1}{m!} 
 \int_0^\ell dx \,  
\Big\{\,
F_m^{(+)}(x) \, \mathcal{H}_+^{(m)}(x)
+
F_m^{(-)}(x) \, \mathcal{H}_-^{(m)}(x) 
\, \Big\}  \,
\ee
where we have introduced the operators
\be
\label{Tplusminus op}
\mathcal{H}_\pm^{(m)} 
\equiv
\sum_{k = 0}^{m}  \binom{m}{k}  \left( -\frac{1}{2}\,\partial_x \right)^{m-k} \Psi^{(k)}_\pm
\ee
and the corresponding weight functions as
 \be
 \label{F_pm_def}
 \begin{array}{l}
 \displaystyle
F_m^{(+)}(x) 
\,\equiv\,
\delta_{m,0} \, t_0(x)  + 2 \, s^m \sum_{r=1}^{\infty}  r^m \cos(r q_F s)  \, t_r(x) \,,
\\
 \displaystyle
F_m^{(-)}(x) 
\,\equiv\,
2 \, s^m\sum_{r=1}^{\infty}  r^m \sin(r q_F s)  \, t_r(x) \,.
\end{array}
\ee
Notice that the $F_m^{(\pm)}(x)$ contribute at order $O(s^m)$
and that the operators $\mathcal{H}_\pm^{(m)}$ are combinations of
$\partial^a \psi^\dagger \partial^b \psi$ where $a+b=m$.

The results obtained in section\;\ref{sec:arb-filling} correspond to the terms with $m=0$ and $m=1$ in (\ref{H high der m-sum}).
Indeed, the leading term has $m=0$ and for the corresponding operators (\ref{Tplusminus op}) 
one finds $\mathcal{H}_+^{(0)} =\Psi^{(0)}_+ $ and $\mathcal{H}_-^{(0)} =0$.
Furthermore, $F_0^{(+)}(x)  = F_0(x) $ defined in (\ref{F_01 def}). 
Thus, the $m=0$ term of (\ref{H high der m-sum}) provides the first integral in (\ref{H_A F01}).
As for the $m=1$ term, by employing the definitions (\ref{Psi's def})
we find that the corresponding operators (\ref{Tplusminus op}) are
$\mathcal{H}_+^{(1)} =0 $ (this operator vanishes because $\Psi^{(1)}_+ = \tfrac{1}{2} \partial_x \Psi^{(0)}_+$) 
and $\mathcal{H}_-^{(1)} =\Psi^{(1)}_- = T_{00}$.
Then, since $F_1^{(-)}(x)  = F_1(x) $ introduced in (\ref{F_01 def}),
we conclude that the $m=1$ term of (\ref{H high der m-sum}) can be written as
the second integral  in (\ref{H_A F01}). 

The $O(s^2)$ and $O(s^3)$ contributions correspond respectively to the $m=2$ and $m=3$
terms in the sum (\ref{H high der m-sum}).
For $m=2$ the operators (\ref{Tplusminus op}) become 
$\mathcal{H}_+^{(2)}= \Psi^{(2)}_+  - \tfrac{1}{2}\,\partial_x \Psi^{(1)}_+ $ and $\mathcal{H}_-^{(2)}=0$.
The latter identity is equivalent to $\Psi^{(2)}_- =  \partial_x \Psi^{(1)}_- = \partial_x T_{00}$.
The contribution of the third derivatives of the fields to the entanglement Hamiltonian (\ref{H high der m-sum})
is due to the $m=3$ term. In this case the operators (\ref{Tplusminus op}) read
$\mathcal{H}_+^{(3)} =\tfrac{1}{8}(  8\Psi^{(3)}_+ - 12\,\partial_x \Psi^{(2)}_+ + 4\,\partial_x^2 \Psi^{(1)}_+ ) =0$ and 
$\mathcal{H}_-^{(3)} =\Psi^{(3)}_- - \tfrac{3}{4}\,\partial_x \Psi^{(2)}_-$.
It would be interesting to further explore the structure of the operators $\mathcal{H}_\pm^{(m)}$.
We expect that $\mathcal{H}_+^{(2p+1)} = \mathcal{H}_-^{(2p)} =0$ for any $p \ge 0$.

At half filling the functions (\ref{F_pm_def}) simplify significantly; indeed they reduce to
 \be
  \label{F_pm half-filling}
F_m^{(+)}(x) = 0 \, ,
\;\; \qquad \;\;
F_m^{(-)}(x) =
2 \, s^m \sum_{p=0}^{\infty}  (-1)^p \,(2p+1)^m  \, t_{2p+1}\,.
\ee
Furthermore, for $m=3$ we can perform the sum in (\ref{F_pm half-filling}) analytically
by following a procedure similar to the one described in section\;\ref{sec:half-filling infinite}.
This leads us to write the expansion  of the entanglement Hamiltonian (\ref{H high der m-sum}) at half filling
restricted to the terms up to the third derivatives included as follows
\be
\label{H_A eps2 half-filled}
\mathcal{H} 
\,=\,
2\pi\, \ell \int_0^\ell dx  \,\beta(x)\,
T_{00}
+
2\pi\, \ell \; \frac{s^2}{3!}
\int_0^\ell  dx \, \beta(x) \Big(\,1-6\,\beta(x)^2\Big)
\left(\Psi^{(3)}_- - \frac{3}{4}\,\partial_x \Psi^{(2)}_- \right)
+\dots
\ee
where $\beta(x)$ is the parabola defined in (\ref{conf_beta}). The weight factor in the second integral
varies as $\beta(x)$ near the ends of the interval, but the maximum at $x=1/2$ is replaced by a
plateau.

\end{appendices}

\section*{References}

\providecommand{\newblock}{}

\end{document}